\documentclass{PoS}

\title{Measurement of Longitudinal Single-Spin Asymmetry for W$^\pm$ Production in Polarized Proton+Proton Collisions at STAR}

\ShortTitle{Measurement of $A_L$ for W$^\pm$ boson in Polarized $p+p$ Collisions at STAR}

\author{\speaker{Qing-hua Xu}, for the STAR collaboration\\
      Key Laboratory of Particle Physics and Particle Irradiation (MoE), \\
       Institute of Frontier and Interdisciplinary Science, \\
       Shandong University, Qingdao, Shandong 266237, China\\
      E-mail: \email{xuqh@sdu.edu.cn}}

\abstract{

The sea quark contribution to the nucleon spin is an important piece for a complete understanding of the nucleon spin structure. The production of $W$ bosons in longitudinally polarized $p + p$ collisions at RHIC provides an unique probe for the sea quark polarization, through the parity-violating single-spin asymmetry, $A_L$. At the STAR experiment, $W$ bosons can be effectively detected through the leptonic decay channel $W \to  e \nu $ with the Electromagnetic Calorimeters and Time Projection Chamber at mid-rapidity. The previous STAR measurements of $A_L$ for $W$ boson production from datasets taken in 2011 and 2012 have provided significant constraints on the helicity distribution functions of $\bar u$ and $\bar d$  quarks. In 2013 the STAR experiment collected $p+p$ data with an integrated luminosity of  about 250 pb$^{-1}$ at $\sqrt s$ = 510 GeV with an average beam polarization of about  $56\%$, which is about three times the total integrated luminosity of previous years. The final $A_L$ results from the STAR 2013 data sample are presented and are also combined with previous 2011+2012 results.  The comparison with theoretical expectations suggests a flavor asymmetry with $\Delta \bar{u}(x)$ $>$$\Delta \bar{d}(x)$ for sea quark helicity distributions with $0.05 < x < 0.25$. 
}
 
\FullConference{23rd International Spin Physics Symposium - SPIN2018 -\\
		10-14 September, 2018\\
		Ferrara, Italy}

\begin{document}


The nucleon spin structure in terms of its parton constituents is a fundmental question in QCD and has been studied extensively both in theory and in experiment since the 1980s.  
The quark spin has been found to contribute about 30\% of the nucleon spin and the valence quark helicity distributions are well determined at intermediate $x$ from DIS experiments.
However, the uncertainties on the sea quark helicity distributions are still relatively large~\cite{WenChen2014}.
The production of $W^{+(-)}$ bosons in $p+p$ collisions with one beam longitudinally polarized provides an unique tool to access sea quark and valence quark helicity distributions without the involvement of hadron fragmentation as in the semi-inclusive DIS process~\cite{BS93,Bunce00,Ma17}.
The longitudinal single-spin asymmetry $A_L$ for $W$ boson production in $p+p$ collisions is defined as:
 \begin{equation}
A_L(W)\equiv \frac {\sigma{(p_+p \to WX)}-\sigma{(p_-p \to WX)}}
{\sigma{(p_+p \to WX)}+\sigma{(p_-p \to WX)}},
 \label{eqa:alw}
\end{equation}
where $\sigma(p_\pm p \to WX)$ is the $W$ cross section with a helicity positive/negative proton beam, with subscripts ``$\pm$'' here denoting the helicity.
As the production of $W$ bosons violates parity maximally,  the above $A_L$ equation for $W^\pm$ 
can be rewritten as follows at leading order:
\begin{equation}
A_L^{W^+} = \frac {-\Delta{u(x_1)\bar d(x_2)}+\Delta{\bar
d(x_1)u(x_2)}} {{u(x_1)\bar d(x_2)}+{\bar d(x_1)u(x_2)}},
\label{eqa:alwp}
\end{equation}
\begin{equation}
A_L^{W^-} = \frac {-\Delta{d(x_1)\bar u(x_2)}+\Delta{\bar
u(x_1)d(x_2)}} {{d(x_1)\bar u(x_2)}+{\bar u(x_1)d(x_2)}}.
\label{eqa:alwm}
\end{equation}
$A_L^{W^+}$ approaches $\Delta \bar d/\bar d$ when $y_W \ll 0$, because the valence quark usually carries a much larger momentum fraction than a sea quark, and the $W$ rapidity along the polarized proton beam is defined as positive. 
Similarly, $A_L^{W^-}$ approaches $\Delta \bar u/\bar u$ when $y_W \ll 0$. 
 
In the experiment, $W^\pm$ can be detected via the leptonic channel $W \to  e \nu $ with the kinematics of the decay lepton alone. 
A theoretical framework that describes inclusive lepton production from $W$ boson decay has been developed, so the measurements of $A_L$ versus the lepton rapidity can be compared with theoretical predictions \cite{RHICBOS,CHE}.
The previous data from $W$ decays at RHIC~\cite{Aggarwal:2010vc,Adamczyk:2014xyw, Adare:2010xa, Adare:2015gsd, Adare:2018csm} provide new constraints on sea quark polarizations through the QCD global analyses of experimental data, which even indicate the existence of a flavor asymmetry in the polarization of the light-quark sea in the proton for parton momentum fractions, $0.05 < x < 0.25$~\cite{Nocera:2014gqa}.
In this contribution, we report new measurements of the single-spin asymmetries for $W^\pm$ bosons in longitudinally-polarized $p+p$ collisions at $\sqrt{s}$ = 510\,GeV from the STAR experiment~\cite{STAR13}, which were finalized shortly after the conference. 
The data were recorded in 2013 by STAR and correspond to an integrated luminosity of about 250 pb$^{-1}$ with an average beam polarization of about 56\%.

%
At the STAR experiment, 
the $W \rightarrow e\nu$ events are characterized by an isolated $e^{\pm}$ with a sizable transverse energy, $E_T^e$, deposited in the Electromagnetic Calorimeter, which peaks near half the $W$ mass, which is referred to as the Jacobian peak.
The neutrino from $W$ decay is undetected, which leads to a large missing transverse energy that appears as an imbalance in the $p_T$ sum of all the reconstructed final state particles.
In contrast, the $p_T$ vector is well balanced for background events such as $Z/\gamma^* \to e^+e^-$ and QCD di-jet or multi-jet events.
The key selection cuts for $W$ signals are thus based on the lepton isolation and $p_T$ imbalance features.
The charge separation is done using the Time Projection Chamber (TPC), 
which covers the full azimuth and a pseudorapidity range of  -1.3$ < \eta <$ 1.3.
The Barrel and Endcap Electromagnetic Calorimeters (BEMC and EEMC) cover full azimuth and pseudorapidity ranges of $-1 < \eta < 1$ and 1.1 < $\eta$ < 2.0 respectively.

\begin{figure}[h]
\begin{center}
\includegraphics[width=150mm]{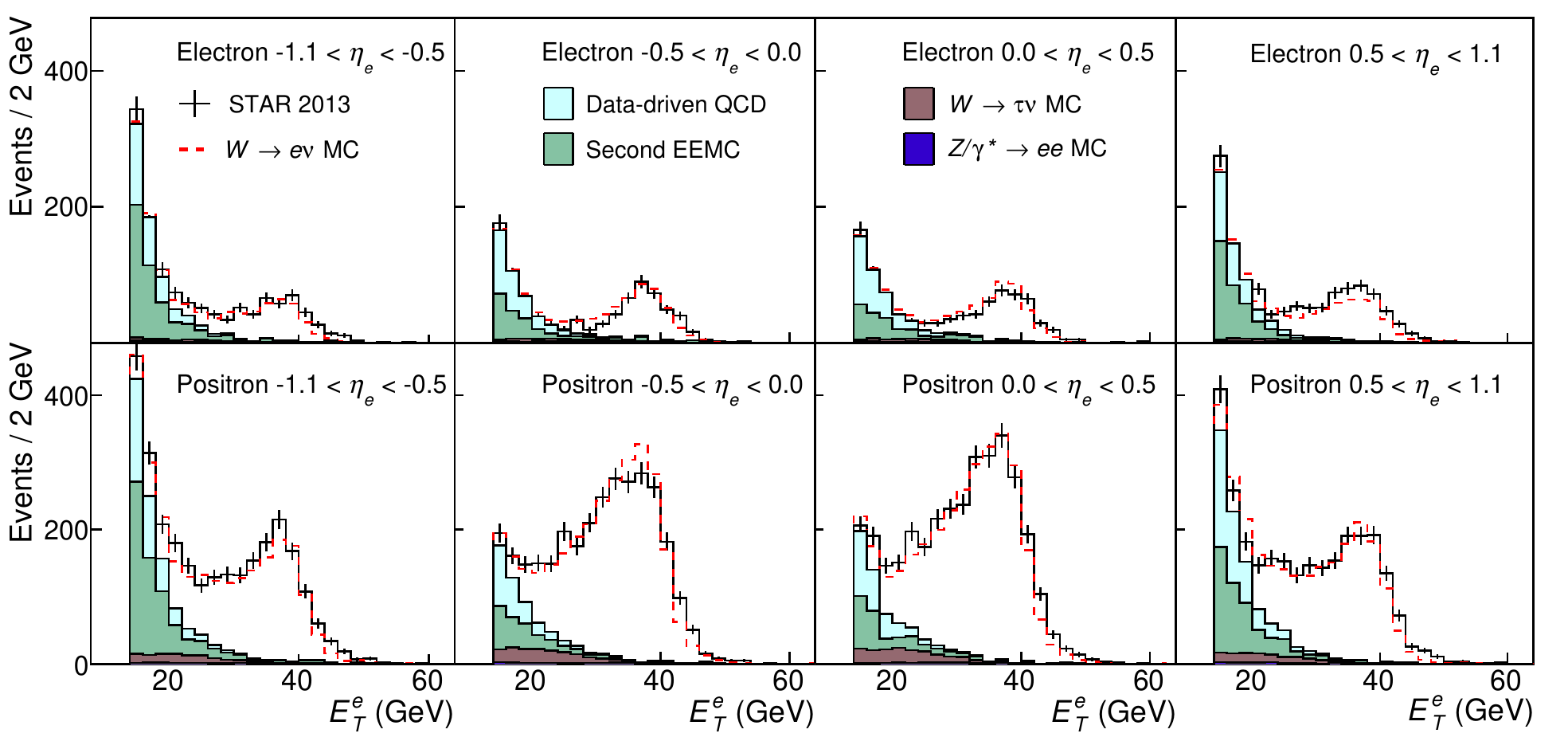}
\caption{
$E_T^e$ distribution for $W^-$ (top) and $W^+$ (bottom) candidates (black), background contributions, and the sum of background and $W\rightarrow e+\nu_e$ Monte Carlo (MC) signals (red dashed) from STAR 2013 data~\cite{STAR13}.
}
\label{fig:Wmass2013}
\end{center}
\end{figure}

First, a charged track with high transverse momentum ($p_T$ > 10 GeV) was chosen, then an EMC energy cluster was reconstructed from $2\times2$ calorimeter towers pointed to by the extrapolated track. The cluster energy $E_T^{2\times 2}$ in these towers is assigned to the candidate electron. 
Then,  further stages of isolation cuts are implemented, with slight differences between BEMC and EEMC regions.
The transverse energy ratio in this 2$\times$ 2 cluster over a surrounding 4$\times$4 cluster is required to be larger than 95\% (96\%) in the BEMC (EEMC) region. 
It is further required that the candidate electron carry a large fraction (larger than 88\% for both BEMC and EEMC) of the energy in the near-side cone with radius $\Delta R=0.7$, where $\Delta R=\sqrt{\Delta \phi ^2+\Delta \eta ^2}$. 
After that, a $p_T$-balance variable is calculated in terms of the $p_T$ sum of the electron candidate and the reconstructed jets outside an isolation cone around the electron track with a radius $\Delta R = 0.7$.
The projection of $p_T$-balance to the direction of the candidate electron (named ``signed $p_T$-balance''), is required to be larger than 14 GeV (20 GeV) for the BEMC (EEMC) region. 
Here, the jet reconstruction used the standard anti-$k_T$ algorithm \cite{anKt}. 
In addition, the total transverse energy in the azimuthally away side is required to be less than 11 GeV to further suppress di-jet backgrounds in the BEMC region when one jet loses a sizable fraction of momentum due to detector effects.
For the EEMC region, another isolation ratio from the energy deposit in the EEMC shower maximum detector (ESMD) is also used ($R_{ESMD}>0.7$)~\cite{STAR13}. 

There are several contributions to the residual background events under the $W$ signal peak, as shown in Fig. \ref{fig:Wmass2013}. 
A QCD di-jet event or $Z/\gamma^* \to e^+e^-$ event can have one of its jets or electrons outside the STAR acceptance and thus pass all the $W$ selection criteria. 
In addition, the $W$ boson can decay to $\tau + \nu$ and $\tau$ can further decay to an electron and a neutrino, though we will not distinguish this feed down contribution.
To estimate the $Z/\gamma^*$ and $\tau$ contributions, 
 the MC events generated using PYTHIA \cite{PYTHIA} through the STAR simulation framework are embedded into STAR zero-bias $p+p$ events and analyzed with the same analysis algorithms.
The QCD background is first estimated using the existing EEMC detector for the uninstrumented acceptance region on the opposite side of the collision point and the rest is estimated using a data-driven method~\cite{Adamczyk:2014xyw}. 
In Fig. \ref{fig:Wmass2013}, the $E_T$ distributions for $W^+$ and $W^-$ are shown for different pseudorapidity intervals covered by the BEMC, where the black histograms are the raw signal and the colored histograms are for different background contributions mentioned above.
The residual background fraction is found to be a few percent and is corrected in the determination of the single-spin asymmetry for $W$ signals, as detailed in Refs.~\cite{Adamczyk:2014xyw,STAR13}.
%
%
%
 %
 \begin{figure}[b]
 \begin{center}
 \includegraphics[width=0.55\textwidth]{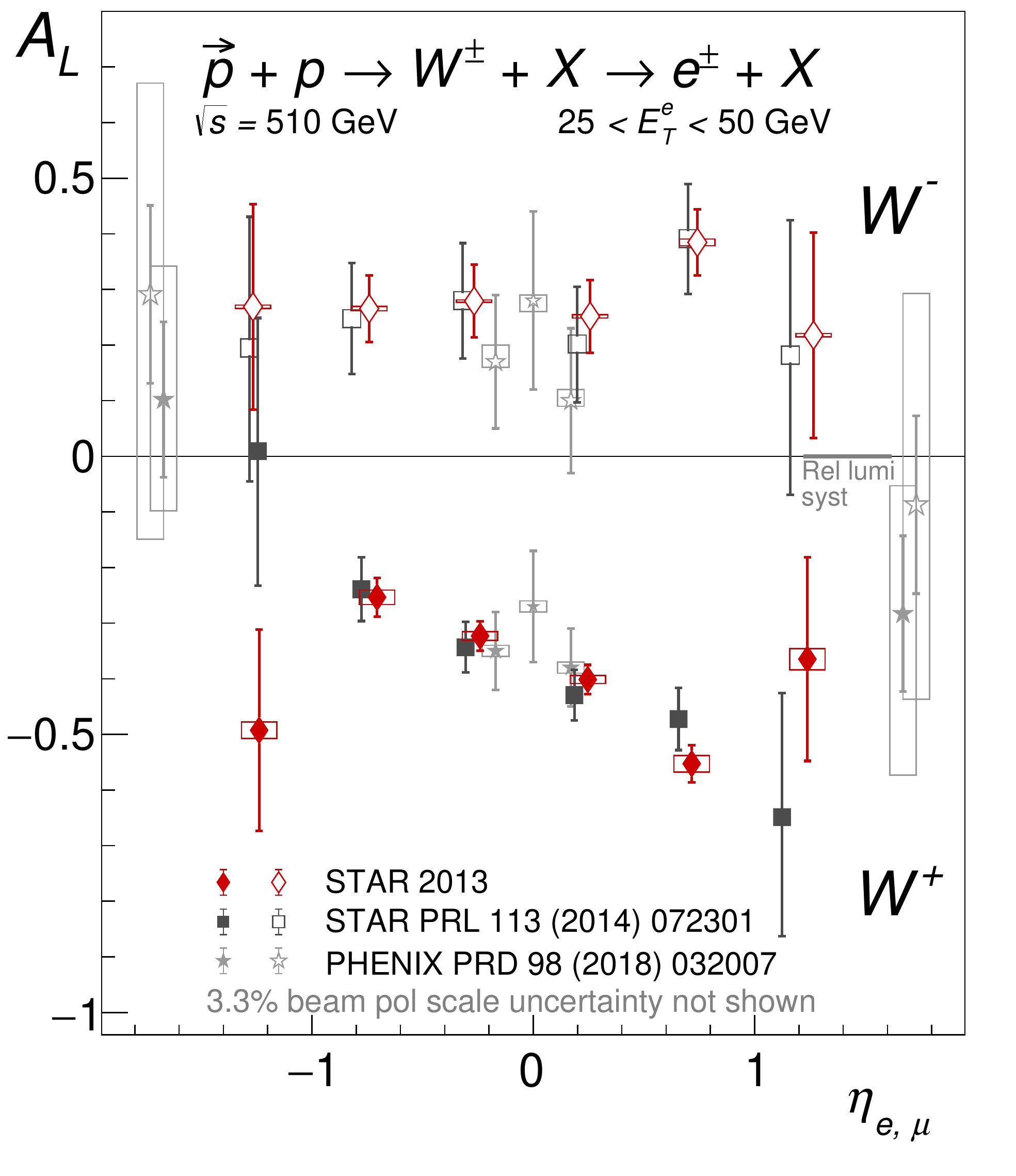} 
 \caption{ Single-spin asymmetry $A_L$ for $W^\pm$ production as a function of $\eta_e$  in $p+p$ collisions at 510 GeV from STAR 2013 data~\cite{STAR13}, in comparison to STAR 2011+2012 data~\cite{Adamczyk:2014xyw} and the PHENIX data~\cite{Adare:2015gsd,Adare:2018csm}.}
 \label{Fig:figure2}
 \vspace*{-5mm}
 \end{center}
\end{figure}

The longitudinal single-spin asymmetry $A_L$ is extracted using the following equation:
\begin{equation}
A_L(W)= \frac{1}{\beta}\frac{1}{P}\frac {N_+/l_+ - N_-/l_-}
{N_+/l_+ + N_-/l_-},
 \label{ALbeta}
\end{equation}
where $\beta$ quantifies the dilution due to residual background, $P$ is the beam polarization, $N_+(N_{-})$ is the $W$ yield obtained in the signal window ($25 < E_T < 50$ GeV) when the polarized beam is helicity positive (negative), and $l_\pm$ are the relative luminosity. 
The relative luminosity factors are calculated independently with high precision using non-$W$ events. 
 
 \begin{figure}[t]
 \begin{center}
 \includegraphics[width=0.60\textwidth]{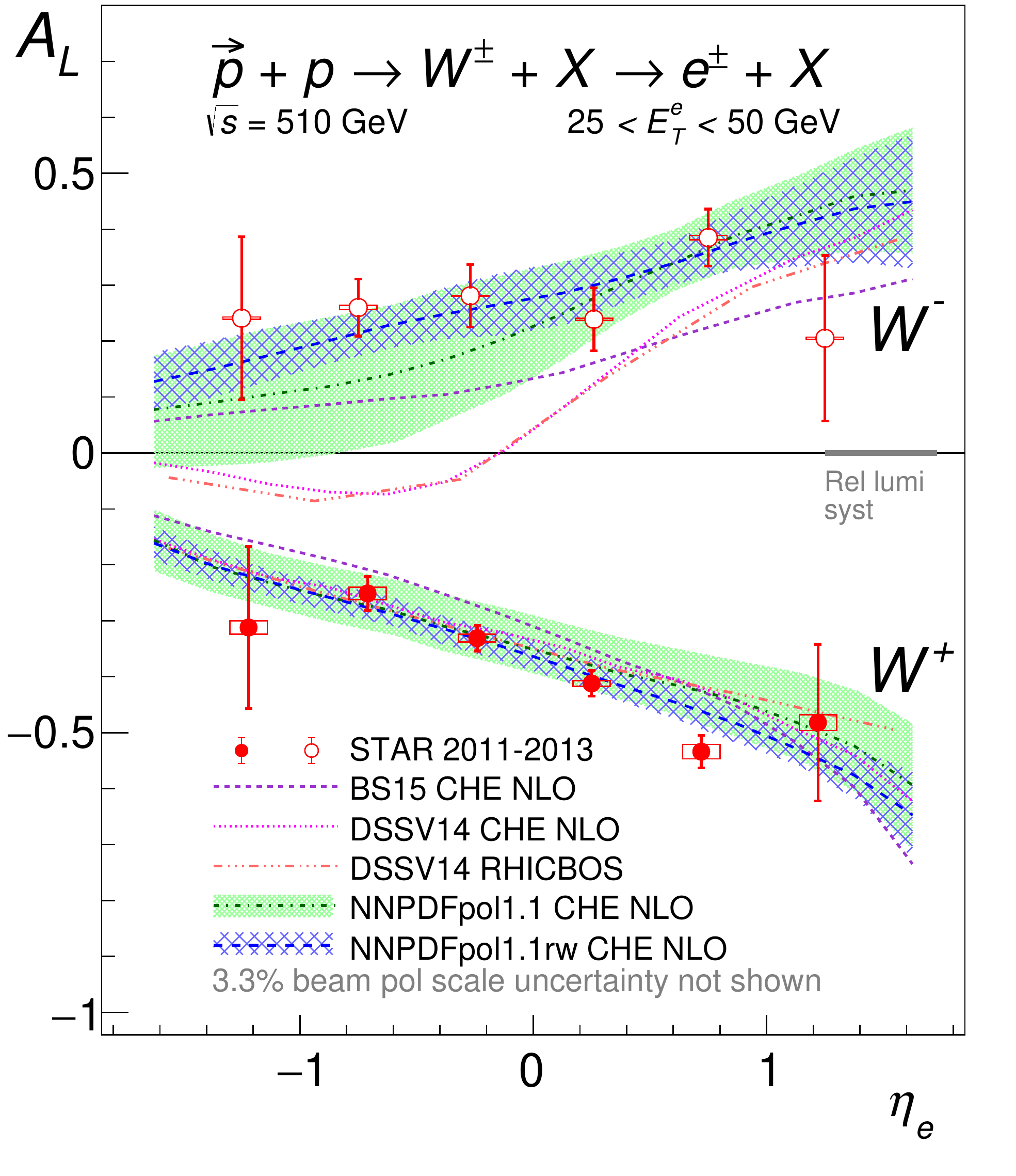} 
 \caption{ Longitudinal single-spin asymmetries, $A_L$, for $W^\pm$ production as a function of $\eta_e$, for the combined 2011+2012 and 2013 STAR data samples~\cite{STAR13} in comparison to theory expectations.}
 \label{Fig:moneyPlotComb}
 \vspace*{-5mm}
 \end{center}
\end{figure}

\begin{figure}[tb]
\centering
\includegraphics[width=8.1cm]{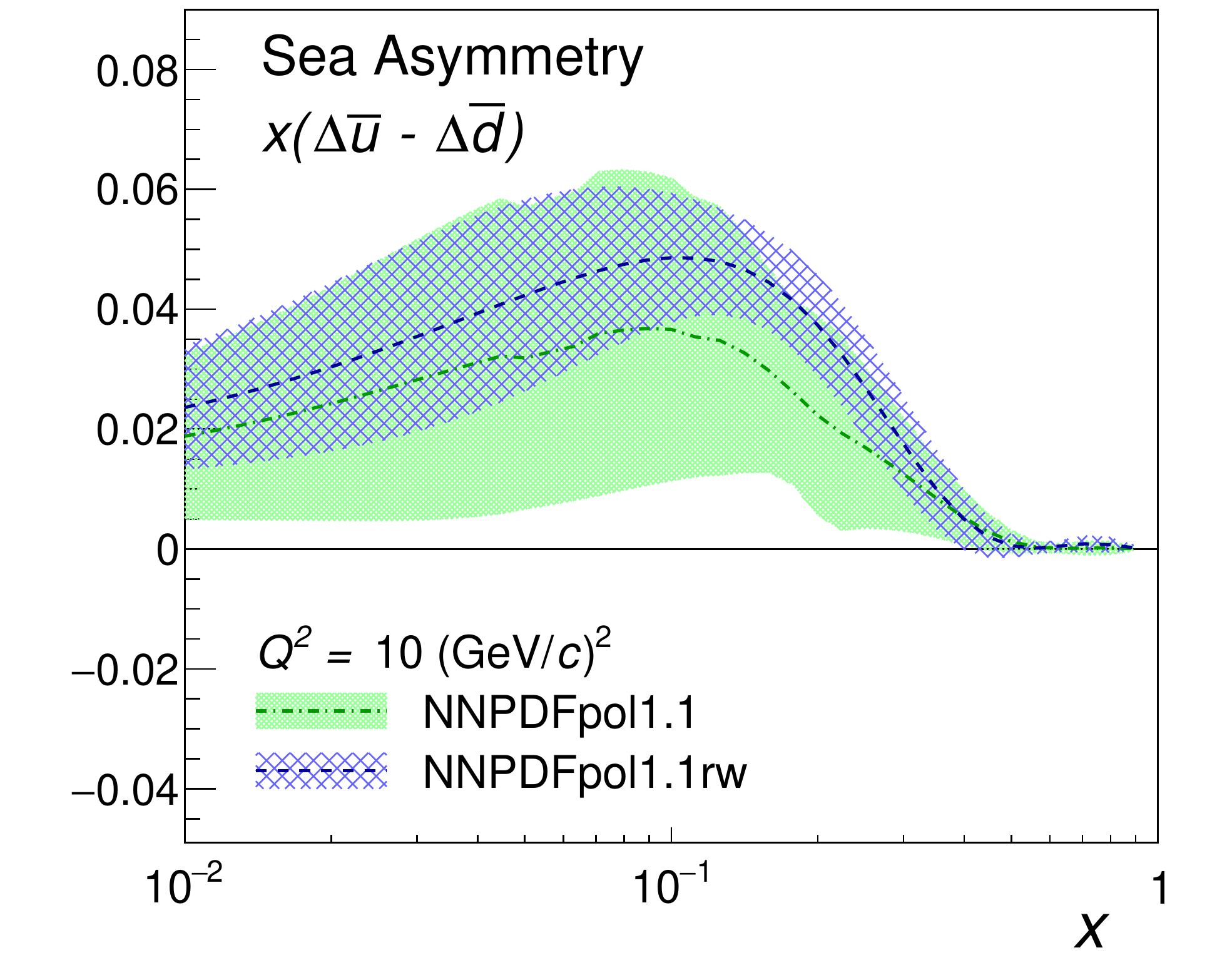}
\caption{\label{fig:PolSea}The difference of the light sea-quark polarizations as a function of $x$ at a scale of $Q^2$ = 10\,(GeV/$c$)$^2$.  The green band shows the NNPDFpol1.1 results~\cite{Nocera:2014gqa} and the blue band shows the corresponding distribution after the 2013 $W^\pm$ data are included by reweighting~\cite{STAR13}.}
\end{figure}

Figure \ref{Fig:figure2} shows the final $A_L$ results for $W^\pm$ boson versus lepton pseudorapidity from the STAR 2013 data sample \cite{STAR13}, in comparison with previous STAR results from the  2011+2012 data \cite{Adamczyk:2014xyw} and the final PHENIX results of $A_L$ for leptons from $W/Z$ decay~\cite{Adare:2015gsd,Adare:2018csm}. 
The systematic uncertainties are indicated by open boxes, which are associated with BEMC and EEMC gain calibrations and the data-driven QCD background.
The STAR 2013 $A_L$ results are consistent with the previous 2011+2012 results, with $40-50\%$ smaller statistical uncertainties. 

The combined STAR data from years 2011-2013 are shown in Fig.~\ref{Fig:moneyPlotComb}, in comparison with theoretical expectations based on the NNPDFpol1.1~\cite{Nocera:2014gqa}, DSSV14~\cite{deFlorian:2014yva} and BS15~\cite{Bourrely:2015} global analyses
under the next-to-leading order \textsc{CHE}~\cite{CHE} and fully resummed \textsc{RHICBOS}~\cite{RHICBOS} frameworks.
The NNPDFpol1.1 expectation includes the STAR $W^\pm$ data from  2011 and 2012~\cite{Adamczyk:2014xyw}, unlike DSSV14 and BS15. 
The STAR 2013 $W$ $A_L$ results have reached unprecedented precision and will significantly advance our understanding of the nucleon spin structure.
To assess their impact, the STAR 2013 data were used in the reweighting procedure of NNPDF global fit based on NNPDFpol1.1 parton densities~\cite{Nocera:2014gqa}. 
The results from the reweighting
are shown in Fig.~\ref{Fig:moneyPlotComb} as the blue hatched bands.
The difference of the helicity distributions of $\bar u$ and $\bar d$ in the proton from the reweighting is shown in Fig.~\ref{fig:PolSea}~\cite{STAR13}.
The new STAR data confirm  the existence of a flavor asymmetry in the polarized quark sea, $\Delta\bar{u}(x)$>$\Delta\bar{d}(x)$, in the range of 0.05 $< x <$ 0.25 at a scale of $Q^2 = 10$\,(GeV/$c$)$^2$.
This is opposite to the flavor asymmetry observed in the unpolarized quark distributions, where $\bar{d}(x)$>$\bar{u}(x)$ over a wide $x$ range has been observed~\cite{WenChen2014}. 

In summary, we report new STAR measurements of the longitudinal single-spin asymmetry for $W^\pm$ bosons produced in polarized proton+proton collisions at $\sqrt{s}$ = 510\,GeV. 
The production of weak bosons provides a unique probe of the sea quark helicity distribution functions of the proton.
The new STAR data, combined with previous STAR results, show a significant preference for $\Delta\bar{u}(x,Q^2) > \Delta\bar{d}(x,Q^2)$ in the fractional momentum range 0.05 $< x <$ 0.25 at $Q^2 \sim 10$\,(GeV/$c$)$^2$.

The author is supported partially by the MoST of China (973 program No. 2014CB845400).

\end{document}